\documentclass[times]{aa}
\usepackage{psfig}
\usepackage{astron}

\begin{document}

\def\eps{\varepsilon}
\def\aap{A\&A}
\def\apj{ApJ}
\def\apjl{ApJL}
\def\mnras{MNRAS}
\def\aj{AJ}
\def\nat{Nature}
\def\azh{Astronomicheskii Zhurnal}
\def\aaps{A\&A Supp.}
\def\st{\sigma_{\rm T}}
\def\th{{\rm th}}
\def\e{{\rm e}}
\def\p{{\rm p}}
\def\me{m_\e}
\def\sync{{\rm s}}
\def\ic{{\rm ic}}
\def\ssc{{\rm ssc}}
\def\cmb{{\rm CMB}}
\def\figsize{0.5}

\title{Synchrotron Self-Comptonized Emission of\\ 
       Low Energy Cosmic Ray Electrons in the Universe:\\ 
       {\Large I) Individual Sources}}

\titlerunning{SSC of Low Energy Cosmic Ray Electrons I}

\author{Torsten A. En{\ss}lin$^1$ \and Rashid A. Sunyaev$^{1,2}$}

\authorrunning{T. A. En{\ss}lin \and R. A. Sunyaev}

\institute{$^1$ Max-Planck-Institut f\"{u}r Astrophysik,
Karl-Schwarzschild-Str.1, Postfach 1317, 85741 Garching, Germany\\
$^2$ Space Research Institute (IKI), Profsoyuznaya 84/32,
Moscow 117810, Russia}

\date{today} 

\abstract{Most of the Universe's populations of low energy cosmic ray
electrons in the energy range of 1-100 MeV still manage to elude from
detection by our instruments, since their synchrotron emission is at
too low frequencies. We investigate a mechanism which can lead to
observable emission of such electron populations: synchrotron-self
Comptonization (SSC). The inverse Compton (IC) scattering can shift an
otherwise unobservable low-frequency 10 kHz-10 MHz photons into
observable radio, infrared (IR) or even more energetic wave-bands. The
resulting emission should be polarized.  We also consider IC
scattering of the cosmic microwave background (CMB) and the cosmic
radio background (CRB).  Electron spectral aging due to synchrotron,
IC and adiabatic losses or gains influences the resulting spectrum.
The predicted radiation spectrum is a strong function of the history
of the source, of the low energy spectrum of relativistic electrons,
and of redshift. It has typically two maxima, and a 
decrement in between at CMB frequencies.  Detection will give a
sensitive probe of the environment of radio galaxies.  We demonstrate
that the fossil remnants of powerful radio galaxies are promising
detection candidates, especially when they are embedded in a dense
intra-cluster medium (ICM).  GHz peaked sources (GPS) have very low
SSC luminosities, which may, however, extend into the X-ray or even
the gamma ray regime. Clusters of galaxies with relativistic electron
populations may be another detectable sources. Fossil radio plasma
released by our own Galaxy could be revealed by its large angular
scale SSC flux. We discuss the expected detectability of these
sources with new and upcoming instruments as LOFAR, GMRT, EVLA, ATA,
ALMA, PLANCK, and HERSCHEL.  
 } \maketitle

\section{Introduction\label{sec:intro}}

\subsection{Exploring a Terra Incognita}

We have detailed knowledge of thermal electrons up to 10 keV in the
wider intergalactic space through their X-ray bremsstrahlung emission
in clusters of galaxies. We know about the existence of extra-galactic
electrons above 1 GeV due to their synchrotron emission in radio
galaxies and cluster radio halos. We have some hints of electron
populations down to 100 MeV due to their IC signatures in the X-ray
and extreme ultraviolet. But the energy range of electrons in
intergalactic space in between 10 keV and 100 MeV is a terra incognita
for our instruments.

Insight into this intermediate energy range would be very important
since the connection between the thermal bulk and the radio emitting
tip of the electron energy distribution could tell us a lot about the
non-thermal processes accelerating particles from low to high
energies.

Our radio observation window is closed below a frequency of $\sim 10$
MHz, due to the ionosphere, and at lower frequencies even due to the
interstellar medium of our own Galaxy.  Therefore, other routes to
probe the low energy electron population in radio sources have been
attempted. A direct way is inverse Compton (IC) scattering of external
photon backgrounds to higher wave-bands. The CMB and IR photons are
scattered by the radio emitting electrons (1-10 GeV) into the X-ray
regime \cite[and
others]{1964GinzburgSyrovatskii,1966ApJ...146..686F,1979ApJ...227..364R,1994ApJ...429..554R,1995MNRAS.273..785H,1995ApJ...449L.149F,1995ApJ...453L..13K,1999ApJ...513L..21F,2000ApJ...534L...7F}. Lower
energy electrons produce extreme ultraviolet photons (0.1-0.3 GeV)
\cite[and
others]{1997Sci...278.1917H,1998AA...330...90E,1998ApJ...494L.177S,1999AA...344..409E,1999ApJ...520..529S,2000ApJ...535..615B}
or even lower energies \cite{1992ApJ...386L...9D}.  A new promising
way seems to be the detection of the IC scattered optical/IR photon
field of the central galaxy. This has only been applied to X-ray data
up to know, giving information about electrons down to 50-100 MeV
\cite{1997A&A...325..898B,1999A&A...342...57B}.  Even lower energy
electrons may already have been seen in quasar jets, if their
large-scale X-ray emission is caused by IC scattering of CMB photons
by a highly relativistic flow
\cite{2000ApJ...544L..23T,2001MNRAS.321L...1C,harris2001}.

Here, an indirect way to peek into this long living, low energy
electron population is proposed. Also not directly observable for us,
these electrons should produce very low frequency (10 kHz - 10 MHz)
synchrotron emission. We propose to detect this indirectly, after IC
scattering of the synchrotron photons by the same electron population,
which produced them.  This SSC emission is at higher frequencies, and
therefore potentially observable by radio, sub-mm, and even higher
frequency telescopes.

The IC optical thickness of most sources is very low, and
therefore for young radio sources the synchrotron emission
dominates over the SSC at radio wavelength. But in old, fossil radio
sources the observable synchrotron emission can be completely absent,
whereas a detectable SSC flux can in principle remain for cosmological
times.

Extragalactic SSC emission caused by higher energy electrons
($\sim$GeV) seems already to be detected.  Harris \& Krawczynski
\cite*{harris2001} list four convincing SSC X-ray detections from
terminal hotspots of radio jets: Cygnus~A
\cite{1994Natur.367..713H,2000ApJ...544L..27W}, 3C~295
\cite{2000ApJ...530L..81H}, 3C~123 \cite{2001MNRAS.323L..17H}, and
3C~263 (Hardcastle et al., in preparation).

The new generation of sensitive telescopes designed to study the tiny
CMB fluctuations (MAP, PLANCK, balloon and ground based CMB
experiments), the new, and planned interferometers in the radio and
sub-mm ranges (LOFAR, GMRT, EVLA, ATA, ALMA), and IR experiments
(HERSCHEL) can open an important window into the low-energy, and often
fossil cosmic ray electron populations, if their their SSC flux can be
detected.  This way we could get unique informations about the origin
and the acceleration mechanisms of low energy relativistic electrons.

\subsection{Clusters of Galaxies}
\begin{figure}[t]
\begin{center}
\psfig{figure=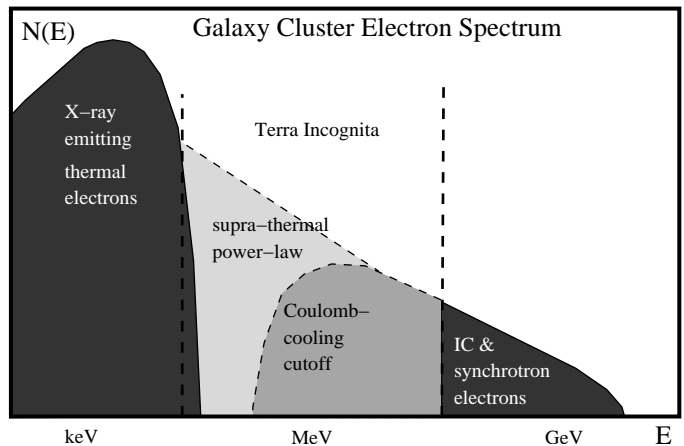,width=\figsize\textwidth,angle=0}
\end{center}
\caption[]{\label{fig:clsketch} Sketch of the possible electron
population in a galaxy cluster containing a radio halo in double
logarithmic units. We have knowledge of the thermal (keV) and of the
high energy (GeV) electron population due to their radiation. The
intermediate region is yet terra incognita for our instruments. The
sketched possible spectral shapes in this region are discussed in the
text.}
\end{figure}

Clusters of galaxies contain thermal electrons with a temperature of
several keV in the ICM. Several clusters also contain electrons with
energies of several GeV, as demonstrated by the presence of cluster
radio halos \cite[for observational
reviews]{1996IAUS..175..333F,1999dtrp.conf....3F}. The origin of these
high energy electrons is still a mystery. Several theories have been
proposed \cite[for a review]{Pune99} such as electron injection by
radio galaxies
\cite{1977ApJ...212....1J,1977ApJ...212..608R,1979ApJ...227..364R,1984A&A...135..141V},
in-situ electron (re-)acceleration by plasma- or shock-waves
\cite{1977ApJ...212....1J,1987A&A...182...21S,1993MNRAS.263...31T,1993ApJ...406..399G,1993ApJ...407L..53R,1996SSRv...75..279V,1997A&A...321...55D,1999ApJ...518..594R,1999ApJ...520..529S,2000ApJ...544..686L,2001MNRAS.320..365B,2000A&A...362..886B,2000ApJ...535..586T},
and secondary particle injection after hadronic interactions of cosmic
ray protons with the background gas
\cite{1980ApJ...239L..93D,1982AJ.....87.1266V,1999APh....12..169B,2000A&A...362..151D}.

The different scenarios lead to very different expectations for the
electron spectrum in the unobserved intermediate energy range. In the
in-situ acceleration scenario, the thermal and ultra-relativistic
electron populations have to be connected by a supra-thermal
population. The exact shape of this population depends on the details
of the acceleration mechanism and the high Coulomb energy losses in
this energy regime
\cite{2000A&A...357...66D,2000ApJ...532L...9B,2001ApJ...557..560P}. In
Fig. \ref{fig:clsketch} one possible realization of such a
supra-thermal population is sketched. But if the electrons are
injected with already high energies into the ICM, e.g. by escape from
radio galaxies, or by hadronic secondary particle production, then the
lower end of the intermediate energy range should be emptied by the
rapid Coulomb energy losses, also sketched in
Fig. \ref{fig:clsketch}. Further, additional spectral features can
exist in this energy range, since the long electron cooling time of
ICM electrons at 100 MeV allows the conservation of the imprints of
earlier electron injection events for cosmological times
\cite{1998ApJ...494L.177S,1999ApJ...520..529S}.

\subsection{Radio Galaxies}
The activity phases of individual radio galaxies, as they appear to
our instruments, are short-lived phenomena compared to cosmological
time-scales. After a few ten million years the central engine
in the galactic center ceases to power the radio cocoon (`radio lobe'
in the radio astronomical jargon), and the
observable radio emission vanishes rapidly
\cite{1987MNRAS.225....1A,1994A&A...285...27K,1998MNRAS.298.1113V,1998MNRAS.297L..86S}.
The lifetime of very low frequency emission (0.01-10 MHz) from radio
galaxies should be much longer than at higher frequencies, due to the
strong energy-dependence of the radiative lifetimes of
ultra-relativistic electrons.

\begin{figure}[t]
\begin{center}
\psfig{figure=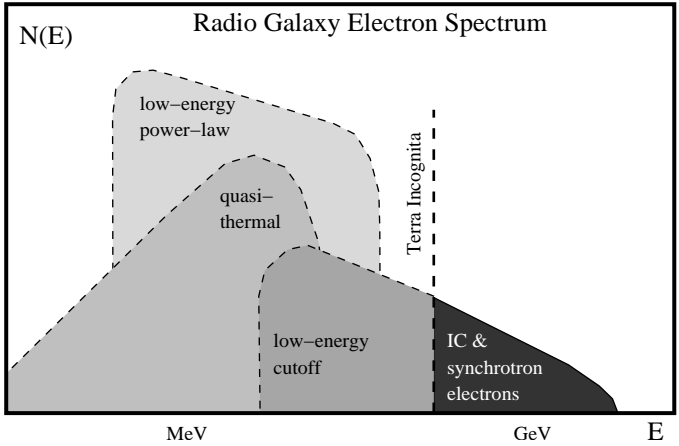,width=\figsize\textwidth,angle=0}
\end{center}
\caption[]{\label{fig:rgsketch} Sketch of the electron population in a
radio galaxy in double logarithmic units. Several spectral features
in the unobserved low energy range are shown. These are a low energy
cutoff to the extrapolated observed spectrum and two possible
distributions which are not usually discussed in the literature: a
quasi thermal component and a separate low energy power law
distribution.}
\end{figure}

The electron energy range below 100 MeV is practically unexplored
yet. These electron have very long lifetimes, and they can be present
a long time after the initial radio galaxy has stopped its activity
\cite{2000A&A...360..417E}, since both their IC radiation and very low
frequency synchrotron emission do not cool them efficiently.  Coulomb
losses are also expected to be small, because of the very low
density a possible thermal component must have in order to be
consistent with the absence of the Faraday depolarization effect in
radio cocoons \cite{1995ApJ...449L.149F}.

Some possible low-energy electron spectra of radio galaxies are
sketched in Fig.  \ref{fig:rgsketch}. It could be that the power-law
radio electron spectrum continues into the unobserved energy range,
and then just has a cutoff. Such a low energy cutoff at an energy
corresponding to an emission frequency of 10 MHz is usually assumed in
the radio astronomical literature in order to calculate equipartition
energy densities from the observed radio emission. This traditional
cutoff at the frequency of the observational frontier gives
equipartition energy densities which are within the order of magnitude
range of reasonable values expected on other grounds. But it has to be
noted that a systematic comparison of equipartition pressure of radio
galaxies with their surrounding ICM pressure indicates that the former
might be too low by a factor between 5 and 10
\cite{1990A&A...233..325F,1992A&A...265....9F,1995A&A...298..699F}. If,
in addition to this pressure discrepancy, radio galaxies are in fact
over-pressured compared to their environments, as many source
evolution scenarios require \cite[for
example]{1989ApJ...345L..21B,1997MNRAS.286..215K,1997MNRAS.292..723K,1999AJ....117..677B},
then there is plenty of space for a large low-energy electron
population and possibly protons, too.

Such an electron population could have a relativistic Maxwell-Boltzmann
distribution, or a separate power-law from an earlier particle
acceleration phase, e.g. within the radio jets of the radio
galaxy. These possibilities are sketched in Fig. \ref{fig:rgsketch},
and many others are possible since our knowledge about the dissipation
processes of relativistic flows in jets are very limited. If mixing of
the radio plasma with the ambient medium occurs, even a
non-relativistic thermal population could be built up with time within
the radio plasma volume.

\begin{figure}[t]
\begin{center}
\psfig{figure=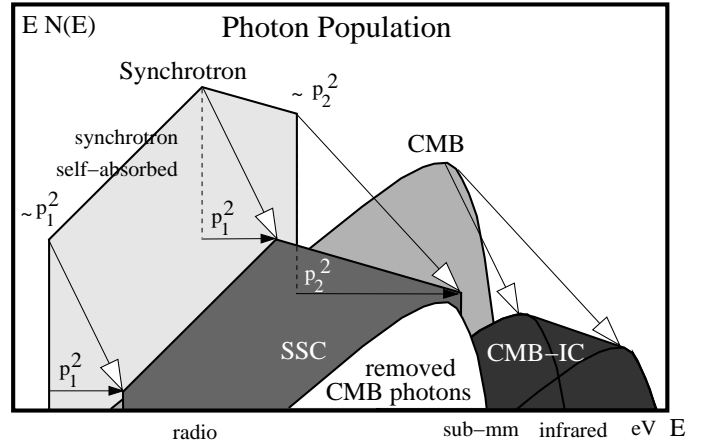,width=0.5\textwidth,angle=0}
\end{center}
\caption[]{\label{fig:sscsketch} Sketch of the SSC and CMB-IC
processes caused by a power-law electron momentum spectrum from
$p_1$ to $p_2$. Shown are the synchrotron and CMB target photon
distributions (the CRB is neglected for
graphical clarity),  the resulting IC scattered
SSC and CMB-IC spectra, and the initial spectrum of the removed,
up-scattered CMB photons. The latter will be recognized as negative
flux by any differentially measuring CMB experiment. The arrows
indicate how the corresponding features in the target and scattered
photon spectra are related to the electron lower and upper cutoff
momenta. For further explanations see the text.}
\end{figure}

\subsection{Radio Ghosts\label{sec:radioghosts}}

Powerful radio galaxies were a hundred times more abundant at
higher redshifts. Therefore their fossils must be a ubiquitous
ingredient of the present Universe, {\it the radio ghosts}
\cite{Ringberg99}. Some of radio ghost's possibly observable
consequences would be their IC signature on the CMB due to their
relativistic electron populations \cite{2000A&A...360..417E}, and the
scattering and isotropization of ultra-high energy comic rays by
their magnetic fields \cite{2001APh....16...47M}.

Indications that radio ghosts indeed exist are the detections of
cavities in the X-ray emitting intra-cluster gas of the Perseus
cluster \cite{1993MNRAS.264L..25B}, the Cygnus-A cluster
\cite{1994MNRAS.270..173C,2000MNRAS.318L..65F}, the Hydra-A cluster
\cite{2000ApJ...534L.135M}, Abell 2597 \cite{McNamara2000Paris}, Abell
4059 \cite{1998ApJ...496..728H,heinz2001}, Abell 2199
\cite{fabian2001moriond}, close to M84 in the Virgo Cluster
\cite{2001ApJ...547L.107F} and RBS797 \cite{schindler2001}.  Whereas
in most of these cases the cavities are obviously the radio cocoons of
a radio galaxy at the cluster center, in some clusters also cavities
without detectable radio emission were found, in Perseus and in Abell
2597 even close to cavities with detected radio emission. This fits
nicely into the picture that the radio plasma outflows of radio
galaxies fill subvolumes of the IGM, which are separated from the
X-ray gas. It shows further that these cavities stay intact even after
the observable radio emission has ceased, as one expects for the radio
ghosts.

Further evidence of the existence of radio ghosts is given by the so
called {\it cluster radio relics}. The properties of these sources can
be understood in the context of cluster merger shock waves compressing
fossil radio plasma clouds, and thereby re-ignite their observable
synchrotron emission by the shock amplification of magnetic fields
and electron energies \cite{1998AA...332..395E}. The agreement between
observed and theoretical spectra
\cite{2001A&A...366...26E,2001AJ....122.1172S}, morphologies, and
polarization characteristics \cite{ensslinbrueggen01} supports this
scenario and the existence of radio ghosts.

We consider below the SSC and CMB-IC spectra of radio ghosts at
different redshifts because of the strong dependence of the high
energy tail of the electron distribution on the CMB energy density.

\subsection{Structure of the Paper}

We give a brief overview about the basic physics in
Sect. \ref{sec:physics}. The details of the formalism  used can
be found in the Appendixes. In Appendix \ref{sec:ecool} the radiative
cooling of an electron population is investigated, in Appendix
\ref{sec:sync} synchrotron emission and absorption are described, and
Appendix \ref{sec:IC} deals with the IC and SSC processes.

This formalism is applied to study the spectral evolution of the SSC
emission, and the IC scattered CMB and CRB\footnote{The cosmic
radio background (CRB) is mostly a superposition of the radiation from
all radio galaxies. Here, it is assumed to have a power-law brightness
spectrum which extends from 1 MHz \cite[for estimates of this
cutoff]{1977MNRAS.180..429S,1996APh.....6...45P} to 1 GHz with a
spectral index of 0.75 and a brightness temperature of 30 Kelvin at
178 MHz \cite{1967MNRAS.136..219B,clark1970}. Our resulting IC
scattered spectra depend only weakly on the uncertainties in the
assumed spectral cutoffs of the CRB, which we modeled only very
crudely. The reason for this is that at the high frequency end the CMB
is dominating over the CRB in any case. And at the low frequency end
either the internal synchrotron photon density dominates over the CRB,
or our sources are optically thick due to synchrotron absorption.}
for a set of characteristic radio sources in Sect. \ref{sec:asscs}:
powerful radio galaxies, like Cygnus A (Sect. \ref{sec:CygA}), GPS
(Sect. \ref{sec:GPS}), giant radio galaxies (Sect. \ref{sec:grg}), the
fossil radio plasma possibly released by our own galaxy
(Sect. \ref{sec:ourghost}), and radio halo containing clusters of
galaxies, like the Coma cluster (Sect. \ref{sec:clogl}).  Our main
findings are summarized in Sect. \ref{sec:concl}.

\section{The Basic Physics\label{sec:physics}}
\subsection{Formalism \& Approximations}

We use monochromatic approximations for the calculation of the
synchrotron and the inverse Compton radiation. This gives accurate
results for the power-law regimes, but surely will be a crude
approximation close to spectral cutoffs. But the uncertainties in the
assumed model parameters do not require a higher accuracy in our
illustrative treatment. Our treatment is not affected by the possible
presence of a relativistic proton component in the radio plasma, since
the model parameters were chosen on the basis of the observed radio
flux. Synchrotron absorption of internal and external radiation fields
is taken into account. The effect of the plasma frequency, and also
the Tsytovich-Razin \cite{tsytovich1951,razin1960}, the free-free, and
induced Compton \cite{1971SvA....15..190S} absorption low frequency
cutoffs of the source internal synchrotron radiation field are
neglected. This is justified due to the very low expected thermal gas
density in radio cocoons.  And even in the much denser galaxy clusters
these effects are only important below $\sim 10$ kHz due to the high
cluster temperatures \cite[for the relevant
formulae]{1999asfo.book.....L}. External free-free absorption of the
resulting low frequency emission can affect the observed SSC spectra
of radio plasma embedded in the interstellar medium of galaxies, as
the GPS sources. But since we calculate the intrinsic source spectrum,
we do not correct for this.

The details of the calculations are laid down in the Appendix.   Similar and more sophisticated formalisms were
derived and applied by many other authors \cite[for
example]{1977ApJ...216..244M,1979A&A....76..306G,1985ApJ...298..128B}.
But several details of the calculation presented in the Appendix
are not -- to our knowledge -- yet reported in the literature:
e.g. formulae for the exact number density (Eq. \ref{eq:nsync}) and total
number (Eq. \ref{eq:Nsync}) of synchrotron photons within a
spherically homogeneous source, the generality of having several
electron populations which can be located in regions with differing
magnetic field strengths, useful asymptotic correct approximations of
the external IC and SSC fluxes as a function of optical depth, and the
inclusion of the IC decrement in external radiation fields into such a
theoretical framework.

\subsection{The Spectral Shape}

In the following we give a qualitative discussion of the properties of
the processes  involved. This should clarify, for example, how our
model spectra would change if different parameters were assumed.
\begin{figure}[t]
\begin{center}
\psfig{figure=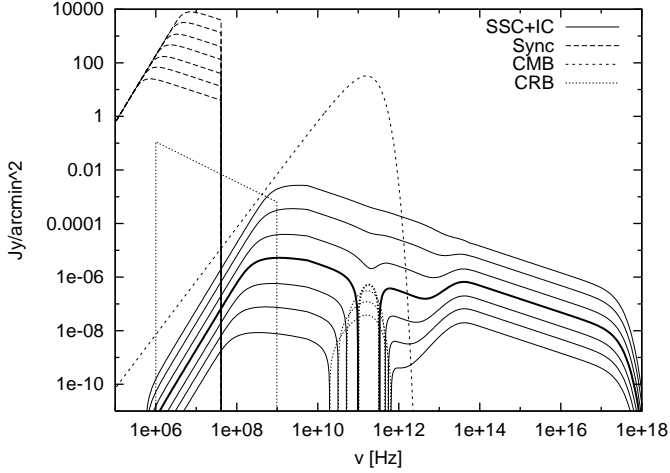,width=\figsize \textwidth,angle=0}
\end{center}
\caption[]{\label{fig:toymodelNorm} Central surface brightness of our
example radio cocoon (see Sect. \ref{sec:example}). The synchrotron
(long-dashed) and SSC+IC spectra (solid) are shown for power-law
normalization parameter $C = 3\cdot 10^{-8},\, 10^{-7},\, 3\cdot
10^{-7},\, 10^{-6},\, 3\cdot 10^{-6},\, 10^{-5},\, \mbox{and}\, 3\cdot
10^{-5} \,{\rm cm^{-3}}$ from bottom to top. The thick lines mark
the case with our reference model with normalization $C= 10^{-6}
\,{\rm cm^{-3}}$. In spectral regions, where the SSC+IC processes
lead to a reduction of the sky brightness below the CMB brightness,
the absolute value of the (negative) SSC+IC surface brightness is
plotted by a dotted line.  The short-dashed line is the CMB
spectrum. The dotted power-law line at frequencies below 1 GHz is the
cosmic radio background (CRB). [Reference model parameters:
$R = 100$ kpc, $B = 5 \,\mu$G, $p_1 = 10$, $p_2 = 10^3$, $s = 2$, $C =
10^{-6}\,{\rm cm^{-3}}$].}
\end{figure}

For the moment, we assume  a power-law (in momentum) relativistic
electron distribution (per volume) with normalization $C$, spectral
index $s$ and lower and upper cutoffs $p_1$ and $p_2$:
\begin{equation}
\label{eq:powerlaw}
f(p)\, dp\,dV = C\,p^{-s}\,\Theta(p-p_{1})\,\Theta(p_{2}-p) \,dp\,dV
\end{equation}
$\Theta(x)$ is the Heaviside step-function, so that for dimensionless
electron momenta $p = P_\e /(m_\e \,c)$ ($P_\e$ is the electron
momentum) outside $p_{1}<p<p_{2}$ the electron spectrum vanishes.
We decided to use the dimensionless momentum $p$ rather than the
canonical $\gamma$ in order to parameterize our electron spectra, since
particle acceleration theories usually predict power law spectra in
momentum rather than in kinetic energy, and the adiabatic losses and
gains are also best described in momentum. In the ultra-relativistic
limit $p$ and $\gamma$ are identical.

If located in a magnetic field with strength $B$, these electrons
produce a synchrotron photon population between the frequencies
\begin{equation}
\nu_{\sync ,1/2} = 3\, e\, B \,p_{1/2}^2/(2\,\pi\,\me\,c) =
\Lambda\,B\,p_{1/2}^2\,,
\end{equation}
where the term $\Lambda = 3\, e/(2\,\pi\,\me\,c)$ is introduced for
convenience. As sketched in Fig. \ref{fig:sscsketch} the synchrotron
spectrum consists (with increasing frequency) of an optically
thick (synchrotron self-absorbed) low frequency part, and a
decreasing optically thin high frequency part (for positive $s$). The
internal synchrotron photons, and also the external CMB and CRB
photons are partly scattered by the electron population to higher
energies. This is also displayed in Fig.  \ref{fig:sscsketch}.

\begin{figure}[t]
\begin{center}
\psfig{figure=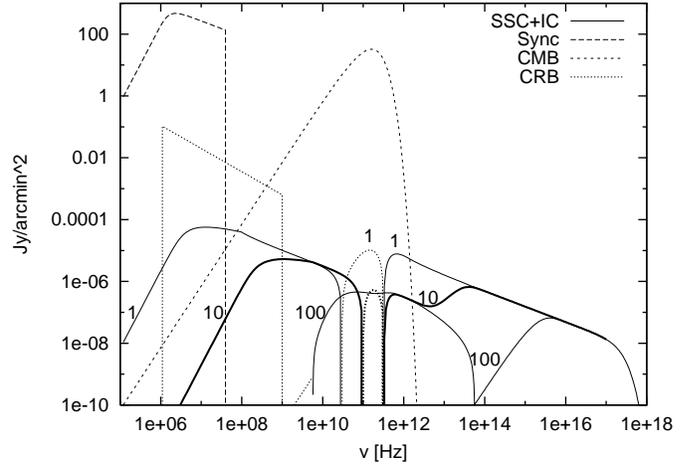,width=\figsize \textwidth,angle=0}
\end{center}
\caption[]{\label{fig:toymodel} The strong dependence of SSC+IC
spectra (solid) on the low energy cutoff in the electron
distribution. The reference model with $p_1 = 10$ is plotted with
thick lines.  The curves are labeled with the electron lower momentum
cutoff $p_1$.  A CMB decrement occurs around 100 GHz in the $p_1 = 1$,
and $p_1 = 10$ models, and below 50 GHz in the $p_1 = 100$
model. The latter is mostly an artifact of our monochromatic
approximation, since it would be partly erased in a calculation in
which the spectral width of the synchrotron and IC emission is taken
into account.  For further details see Fig. \ref{fig:toymodelNorm}
and Sect. \ref{sec:example}.  [Reference model parameters: $R =
100$ kpc, $B = 5 \,\mu$G, $p_1 = 10$, $p_2 = 10^3$, $s = 2$, $C =
10^{-6}\,{\rm cm^{-3}}$].}
\end{figure}

Ultra-relativistic electrons of momentum $p$ scatter photons with
frequency $\nu$ to typically $\nu' = \frac{4}{3}\, p^2\,\nu$. The SSC
spectrum therefore extends from $\nu_{\ssc ,1} =
\frac{4}{3}\,\Lambda\, B \,p_1^4$ to $\nu_{\ssc ,2} =
\frac{4}{3}\,\Lambda\, B \,p_2^4$. If the synchrotron and the SSC
cutoff of a source can be identified (or any other feature in these
spectra caused by a peculiarity in the electron spectrum), the
magnetic field strength of the source is immediately known:
\begin{equation}
B = \frac{4\,\nu_{\sync ,2}^2}{3\,\Lambda\,\nu_{\ssc ,2}}\,.
\end{equation}
The peak of the synchrotron spectrum translates into a similar peak in
the SSC spectrum if the electron number spectrum has itself a
distinctive peak. For steep power-law spectra, as mostly considered in
this work, this is always the lower energy cutoff region. Therefore
the synchrotron peak frequency gets shifted by a factor
$\frac{4}{3}\,p_1^2$ to higher frequencies. This allows to derive the
low energy cutoff of the electron spectrum from observing both, the
synchrotron and the SSC peak. Or vice versa, a change of the cutoff
parameter $p_1$ in one of our models results in a shift of the SSC
peak location, without a significant change at higher frequencies. If
$p_1$ is reduced, the high frequency SSC power law has to be continued
at its low frequency end. The low frequency SSC spectrum below the IC
scattered synchrotron self-absorption peak would increases
correspondingly. A similar statement is true for the peak of the
CMB-IC spectrum. An illustration of these effects is given by
Fig. \ref{fig:toymodel}.

\subsection{Timescales}

The lifetime of particles radiating at a given frequency is significantly
greater for the SSC process than for synchrotron emission. If only
synchrotron and IC cooling are important, the cooling time at momentum
$p$ is
\begin{equation}
t_{\rm cool}(p) = (a_0\,(u_B + u_C)\,p)^{-1}\,,
\end{equation}
where $a_{0} = \frac{4}{3}\, \st/(m_\e\, c)$, and $u_B$ and $u_C$ are
the magnetic and photon (mainly CMB) energy density. Therefore, 
the typical lifetimes at a given frequency $\nu$ are
\begin{equation}
t_{\sync}(\nu) =
\frac{({\nu}/({\Lambda\,B}))^{-\frac{1}{2}}}{a_0\,(u_B + u_C)} \ll
t_{\ssc}(\nu) = \frac{({3\,\nu}/({4\,\Lambda\,B}))^{-\frac{1}{4}}}{a_0\,(u_B + u_C)}\,.
\end{equation}
The ratio of the lifetimes is only weakly dependent on the magnetic
field strength and the frequency itself:
\begin{equation}
\frac{t_\ssc(\nu)}{t_\sync(\nu)} = \left( \frac{4\,\nu}{3\,\Lambda\,B}
\right)^{\frac{1}{4}} = 112 \, \left( \frac{\nu}{\rm
GHz}\right)^{\frac{1}{4}}\,\left( \frac{B}{\mu \rm G}\right)^{-\frac{1}{4}}
\end{equation}
As a rule of thumb, the SSC flux lasts two orders of magnitude longer
than the synchrotron emission, if only synchrotron and IC losses are
important. This should apply for galaxy cluster radio halos and their
descendent electron populations, but usually not for radio galaxies
which are still expanding and therefore have strong adiabatic losses.

\begin{figure}[t]
\begin{center}
\psfig{figure=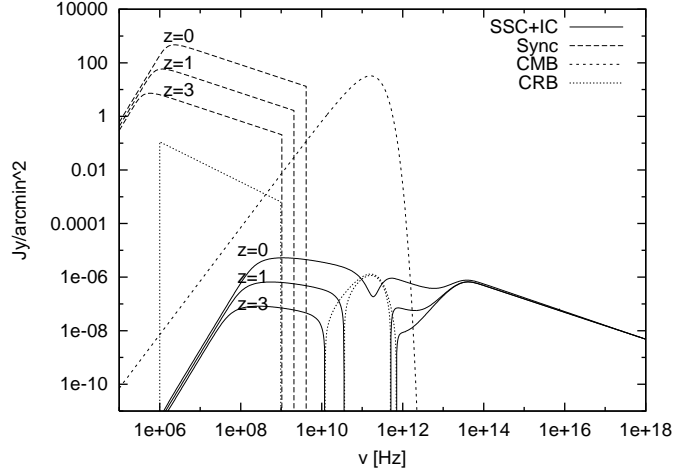,width=\figsize \textwidth,angle=0}
\end{center}
\caption[]{\label{fig:toymodelz} Redshift dependence of the SSC+IC
spectra. Displayed is the central surface brightness of our example
radio cocoon [$R = 100$ kpc, $B = 5 \,\mu$G, $p_1 = 10$, $p_2 =
10^4$ (!), $s = 2$, $C = 10^{-6}\,{\rm cm^{-3}}$] for redshifts
$z=0$, $z=1$, and $z=3$.  For further details see
Fig. \ref{fig:toymodelNorm} and Sect. \ref{sec:example}.  }
\end{figure}

\subsection{Luminosities}

The synchrotron, SSC, and CMB-IC luminosities scale differently with
the source properties. The SSC luminosity depends very sensitively on
the source internal energy density. The synchrotron luminosity also
strongly depends on the source energy density, but to a lower degree
than the SSC luminosity. The CMB-IC luminosity is least
dependent. This is illustrated in Fig. \ref{fig:toymodelNorm}.

Adiabatic compression of the source volume $V$ by a factor $\tilde{C}$
changes the electron spectrum ($f(p)\,dp\,dV$) normalization by $C
\propto \tilde{C}^{({s+2})/{3}}$, and the magnetic field strength by
$B\propto \tilde{C}^{{2}/{3}}$. In the optically thin power-law
regime, the synchrotron luminosity scales like
\begin{equation}
L_\sync(\nu) \propto V\,C\,B^{\frac{s+1}{2}} \propto  \tilde{C}^{\frac{2 s}{3}}
\end{equation}
(Eq. \ref{eq:Lsync2}), the SSC luminosity with
\begin{equation}
L_\ssc(\nu) \propto V^{\frac{4}{3}}\,C^2\,B^{\frac{s+1}{2}} \propto
\tilde{C}^{\frac{3 s+1}{3}}
\end{equation}
(Eq. \ref{eq:LSSC}), and  the CMB-IC (or similar CRB-IC) luminosity
scales like
\begin{equation}
L_{\rm CMB-IC}(\nu) \propto V\,C \propto \tilde{C}^{\frac{s-1}{3}}
\end{equation}
(Eq. \ref{eq:Lic}).  If the compression of a radio plasma cocoon is
due to an environmental pressure change from $P_1$ to $P_2$, e.g. in a
cluster merger shock wave, the adiabatic relation of relativistic
plasma $(P_2/P_1)= \tilde{C}^{{4}/{3}}$ leads to $L_\sync \propto
(P_2/P_1)^{{s}/{2}}$, $L_\ssc \propto (P_2/P_1)^{({3s+1})/{4}}$, and
$L_{\rm CMB-IC} \propto (P_2/P_1)^{({s-1})/{4}}$.  From this, one sees
that synchrotron and SSC emission are strongly, but CMB-IC
luminosity is only weakly affected by compression. The spectral
distortions in the CMB frequency range by the CMB-IC process is mostly
an absorption feature due to the removed, up-scattered photons. Its
strength depends only on the total number of ultra-relativistic
electrons, and is therefore completely independent of the compression
state of the radio plasma \cite{2000A&A...360..417E}. 

\begin{figure}[t]
\begin{center}
\psfig{figure=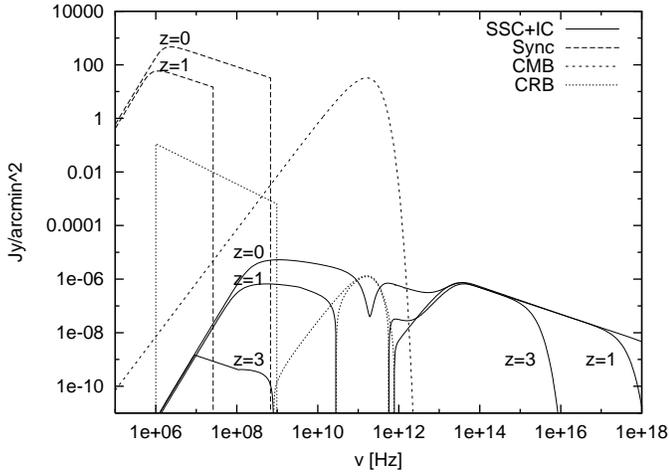,width=\figsize \textwidth,angle=0}
\end{center}
\caption[]{\label{fig:toymodelz100Myr} As Fig. \ref{fig:toymodelz},
but after 100 Myr of radiative cooling. The $z=3$ model's remaining
synchrotron emission is at too low frequencies to be displayed.}
\end{figure}

\subsection{Particular Examples\label{sec:example}}

In order to demonstrate the richness of information contained in the
SSC + CMB-IC spectra we investigate the following example: the radius
of an old radio cocoon is assumed to be 100 kpc, the magnetic field
strength $B = 5 \,\mu$G, the power-law electron spectrum extends from
$p_1 = 10$ to $p_2 = 1000$ with a spectral index $s = 2$ and a
normalization of $C = 10^{-6}\,{\rm cm^{-3}}$.

The spectrum of this cocoon is displayed in
Fig. \ref{fig:toymodelNorm}, together with similar models where the
electron spectrum normalization $C$ is varied (partially to
unrealistic values). Since synchrotron and CMB-IC spectra are
proportional to $C$, but the SSC spectrum is proportional to $C^2$ the
total source spectrum change its shape significantly with a variation
of $C$. A CMB-IC decrement centered on $\sim 100$ GHz appears for
sufficiently low normalizations of the electron spectrum.

In Fig. \ref{fig:toymodel} different models are displayed, where now
the low energy cutoff is changed.  A strong dependence of the
resulting spectra on this cutoff is obvious. With decreasing $p_1$ the
electron number density strongly increases and the CMB-IC decrement
(increment) becomes dominant around 100 GHz (above 1 THz). Also the
SSC emission at low radio frequencies increases with decreasing $p_1$.

In Fig. \ref{fig:toymodelz} the model cocoons (here with $p_2 = 10^4$)
are located at different redshifts ($z= 0,1,3$). The CRB co-moving
photon density is assumed to be reduced to 50\% at $z=1$, and 10\%
($z=3$) compared to its present value. Since the synchrotron and SSC
brightnesses suffer from the cosmological redshift, but the CMB-IC
flux does not, a high redshift source can have a qualitatively very
different SSC + CMB-IC spectrum compared to its low redshift
counterpart. In addition to this, after the shutdown of the radio jets,
these differences are further amplified by the action of radiative
cooling, which is stronger at higher redshifts.  This is illustrated
by Fig. \ref{fig:toymodelz100Myr}, where the spectra of our models are
shown after 100 Myr cooling.

\begin{figure}[t]
\begin{center}
\psfig{figure=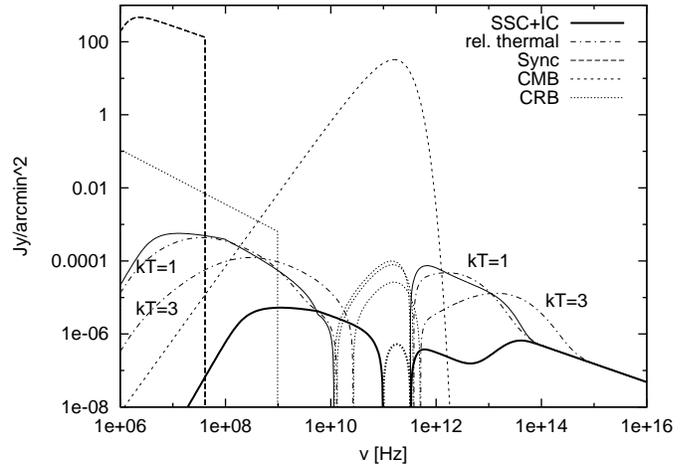,width=\figsize \textwidth,angle=0}
\end{center}
\caption[]{\label{fig:toymodellEPop} The dependence of SSC+IC spectra
in the case of presence of additional components of low energy
relativistic electrons. The thick lines mark the case with our
reference model. 1) The upper solid line is with an additional
power-law component ($p_1 = 1$, $p_2 = 10$, $s = 2$, $C =
10^{-6}\,{\rm cm^{-3}}$ which is a 10 times higher normalization than
at higher energies). 2) The dot-dashed lines are models with added
relativistic thermal populations with the same amount of kinetic
energy as the additional power-law model ($kT_\e = 1\,m_\e\,c^2$ with
$n_{\e,\th} = 8.4\cdot 10^{-6}\,{\rm cm^{-3}}$ and $kT_\e =
3\,m_\e\,c^2$ with $n_{\e,\th} = 2.8\cdot 10^{-6}\,{\rm cm^{-3}}$, see
Eq. \ref{eq:fth}). For further details see
Fig. \ref{fig:toymodelNorm}. [Reference model parameters: $R = 100$
kpc, $B = 5 \,\mu$G, $p_1 = 10$, $p_2 = 10^3$, $s = 2$, $C =
10^{-6}\,{\rm cm^{-3}}$].}
\end{figure}

With Fig. \ref{fig:toymodellEPop} we demonstrate that the combined SSC
and CMB-IC spectra contain enough information in order to discriminate
between models with different energy-rich low energy electron
populations, e.g. power-law and relativistic Maxwell-Boltzmann
distributions (Eq. \ref{eq:fth}) as e.g. sketched in
Fig. \ref{fig:rgsketch}.

\begin{table*}[bth]
\begin{tabular}{|l|r|r|r|l|}
\hline
Instrument & Frequency & Sensitivity & FWHM & {\tt http://} \\
\hline
LOFAR      & 10-300 MHz    &   $\sim$ mJy  	& $20''-1''$ & 
{\tt \footnotesize rsd-www.nrl.navy.mil/7213/lazio/decade\_web}\\
LOFAR core & 10-300 MHz    &   $\sim$4 mJy  	& $50'-3'$ & 
{\tt \footnotesize rsd-www.nrl.navy.mil/7213/lazio/decade\_web}\\
GMRT       & 0.1-1 GHz     &   150-30 $\mu$Jy  	& $20''-2''$ & 
{\tt \footnotesize www.ncra.tifr.res.in/ncra\_hpage/gmrt/gmrt\_spec.html}\\
GMRT core  & 0.1-1 GHz     &   300-60 $\mu$Jy  	& $7'-0.7'$ & 
{\tt \footnotesize www.ncra.tifr.res.in/ncra\_hpage/gmrt/gmrt\_spec.html}\\
ATA   	   & 1-10 GHz 	& 30   $\mu$Jy  	& $75''-9''$ & 
{\tt \footnotesize astron.berkeley.edu/ral/ATAImagerDocs.html}\\
EVLA	   & 1-50 GHz	& 2-20 $\mu$Jy     	& $3'-0.1'$ & 
{\tt \footnotesize www.aoc.nrao.edu/doc/vla/EVLA}\\
ALMA	   & 35-850 GHz	& 2-450  $\mu$Jy     	& $10''-0.5''$ &
{\tt \footnotesize www.eso.org:8082/info/sensitivities/index.html}\\
PLANCK LFI & 30-100 GHz 	& 13-27 mJy  	& $33'-10'$ &
{\tt \footnotesize astro.estec.esa.nl/Planck/technical/payl/node6.html}\\
PLANCK HFI & 0.1-0.9 THz 	& 10-40 mJy  	& $11'-5'$ &
{\tt \footnotesize astro.estec.esa.nl/Planck/technical/payl/node7.html}\\
HERSCHEL   & 0.4-4 THz	& $\sim$ mJy   & $\sim 10''$  & 
{\tt \footnotesize astro.estec.esa.nl/astrogen/first/experiments\_frame.html}\\
\hline
\end{tabular}
\caption[]{\label{tab:sensitivity}Technical specifications of planed
instruments (1st row), as advertised on their Internet pages (5st
row): observing frequencies (2st row), 1-sigma point source
sensitivity (3rd row), and beam FWHM (4th row).  Sensitivities are given
for 1 h observation, except in the case of the PLANCK instruments,
where they are given for the completed PLANCK survey. For the EVLA and
ALMA the compact array configurations were assumed. Note, that most of
these instruments are still under design, so that the values given
here may become obsolete.}
\end{table*}

\subsection{Polarization}

IC scattering of an anisotropic seed photon field leads to some
polarization in the scattered radiation for small electron Lorentz
factors (see e.g. Begelman et al. \cite*{1987ApJ...322..650B}). In our spherically
symmetric model the photon field has a radial anisotropy at off-center
locations. Therefore the source should exhibit an SSC polarization
pattern, which is strongest at the source edge and has electric
vectors aligned on circles. The source averaged polarization is zero
in our highly symmetric toy model. But in nature, typical sources can
be expected to be elongated. For radio galaxy lobes, elongation by a
factor of two are not unusual, and rather large aspect ratios can be
attributed to head-tail radio galaxies. Thus, we expect radio galaxies
to exhibit a small level of linearly polarized SSC flux at lowest
frequencies due to source elongation, if a substantial only mildly
relativistic electron population is present in the source.

Further, the intrinsic synchrotron polarization of sources with
large-scale ordered magnetic fields is 
conserved to a large fraction in the IC process
\cite{1973A&A....23....9B,1994MNRAS.268..451C}. Since many radio
galaxies exhibit polarization, fossil radio galaxies should often have
an intrinsically polarized SSC flux of up to 40~\% polarization at
all frequencies.

Environmental influences on the radio plasma, like shear flows and
shock compression, can increase both polarization effects by
increasing the source elongation and by aligning the internal magnetic
fields \cite[for the impact of shock waves]{ensslinbrueggen01}.

Very peculiar polarization properties can arise around 100 GHz for
sources, where the unpolarized CMB is diminished by IC scattering, but
polarized SSC emission contributes to this spectral range. If the
CMB-IC decrement exceeds the SSC luminosity a polarized decrement in
the CMB sky can be found at these frequencies. This can be of
importance for future high precision CMB polarization measurements.

\section{Upcoming Instruments\label{sec:newinstruments}}

Several telescopes are under development which may be suitable to
search for SSC and CMB-IC signatures of low energy cosmic ray
electrons. We list in Tab. \ref{tab:sensitivity} approximate values of
expected point source sensitivities $F_{\nu}$ of several planed
instruments. If a source with surface brightness $S_\nu$ and angular
area $A_{\rm s}$ is smaller or comparable to the instrument beam with
area $A_{\rm b}$ then the full sensitivity can directly be compared to
the source luminosity in order to estimate the significance $\sigma$
of the expected detection: $\sigma \approx S_\nu \,A_{\rm s}/F_{\nu}$.
If the source appears extended the significance is lower. For a
single-dish and phased array telescopes the significance is
approximately given by $\sigma \approx S_\nu \,(A_{\rm s}\,A_{\rm
b})^{1/2}/F_{\nu}$, since the signal of the $N = A_{\rm s}/A_{\rm b}$
individual beams should be added like independent measurements
yielding $\sigma_{\rm total} \approx \sigma_{\rm beam} \,
N^{1/2}$. For interferometers, the detected signal of an extended
source scales with (the square root of) the fraction of the telescope
baselines for which the source is still unresolved. This depends
strongly on the individual array design and we therefore do not
attempt to estimate it. Often, the interferometers elements are
distributed in a way that many more short baselines (which are
sensitive to large-scale structures) exist than long baselines (which
only see small angular sources). For example it is planed that
$\approx 25$\% of the collecting area of LOFAR is within a virtual
core of 2 km diameter (compared to an instrument diameter of 300
km). Therefore the virtual core of LOFAR is quiet sensitive to
extended flux on angular scales 60 times larger than the full
resolution of the instrument.

\section{Astrophysical SSC Sources\label{sec:asscs}}

In the following models of several extragalactic SSC and CMB-IC
sources are discussed.  The exact definitions of the model parameters
can be found in the Appendix. We give hints on the detectability of
theses sources by the instruments listed in Tab. \ref{tab:sensitivity}
based on rough detection significance estimates as described in
Sect. \ref{sec:newinstruments}.

\subsection{Cygnus A-like Radio Galaxies\label{sec:CygA}}

\begin{figure}[t]
\begin{center}
\psfig{figure=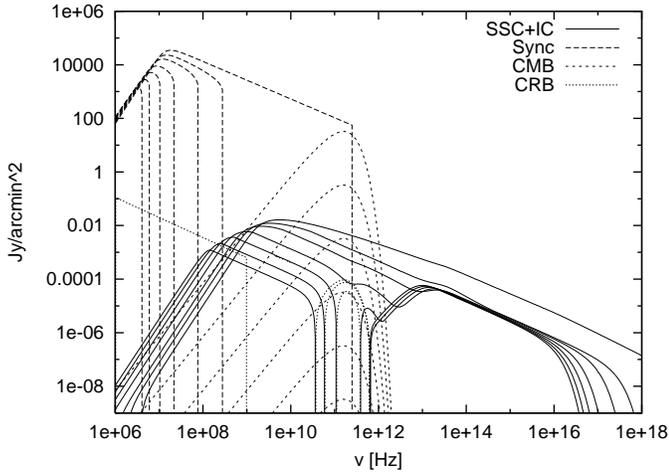,width=\figsize \textwidth,angle=0}
\end{center}
\caption[]{\label{fig:CygAexp} Central surface brightness of our
Cygnus A-like radio cocoon in a cooling and expanding phase.  It is
assumed that the radio plasma supply was shut down in a stage
corresponding to the present Cygnus A stage and the radio cocoon
expanded afterwards during its buoyant rise in the cluster atmosphere.
The synchrotron (long-dashed) and SSC+IC spectra (solid) are shown for
the stages at the jet-power shutdown, and for stages 20, 40, 80, 120,
160, and 200 Myr after this. In spectral regions, where the SSC+IC
processes lead to a reduction of the brightness below the CMB
brightness, the absolute value of the (negative) SSC+IC surface
brightness is plotted by a dotted line.  The top one of the
short-dashed lines is the CMB spectrum. The short-dashed lines
below this is the CMB spectrum multiplied by diminishing factors of
$10^{-2}, 10^{-4}, 10^{-6}, 10^{-8}, \,\mbox{and}\,10^{-10}$ in order
to allow a convenient read off of the relative source brightness
expressed in terms of the CMB brightness at the same frequency.  The
dotted power-law line at frequencies below 1 GHz is the cosmic radio
background (CRB). [Model parameters: $R = 20$ kpc, $B = 35 \,\mu$G,
$p_1 = 10$, $p_2 = 3\cdot 10^4$, $s = 2.4$, $C = 6.63\cdot
10^{-4}\,{\rm cm^{-3}}$, $b = 1.5$, $t_0= 75$ Myr].}
\end{figure}

Our high luminosity radio galaxy is assumed to have spherical lobes
with a diameter of 40 kpc, and a magnetic field strength of
35~$\mu$G. The electron population is a single power-law with spectral
index of $s = 2.4$, which extends from $p_{1} = 10$ to $p_{2} =
3\,\cdot 10^4$ and which is normalized by $C = 6.63\cdot10^{-4}\, {\rm
cm}^{-3}$. This corresponds to an energy content in relativistic
electrons of $4.9\cdot 10^{59}$ erg and in magnetic fields of
$4.8\cdot 10^{58}$ erg. The energy ratio of electronic to magnetic
energy is 10, and therefore by only a factor of a few above what
radio-astronomers call equipartition, for which the energy of
electrons radiating above 10 MHz equals that of the magnetic
fields. The luminosity at 1 GHz is $3.9\cdot10^{34}\,{\rm
erg/Hz/s}$. Cygnus A, for example, has a flux of 214 Jy at 8 GHz
\cite{1971AJ.....76....1S}, as two of our model cocoons if they
are placed at the luminosity distance of 265 Mpc, similar to that of
Cygnus A for $H_0 = 65\,{\rm km/s/Mpc}$ and $q_0 = 0.5$ \cite[for the
redshift]{1997ApJ...488L..15O}. The synchrotron and SSC central
surface brightness of our radio cocoons are shown in
Fig. \ref{fig:CygAexp}.  

\begin{figure}[t]
\begin{center}
\psfig{figure=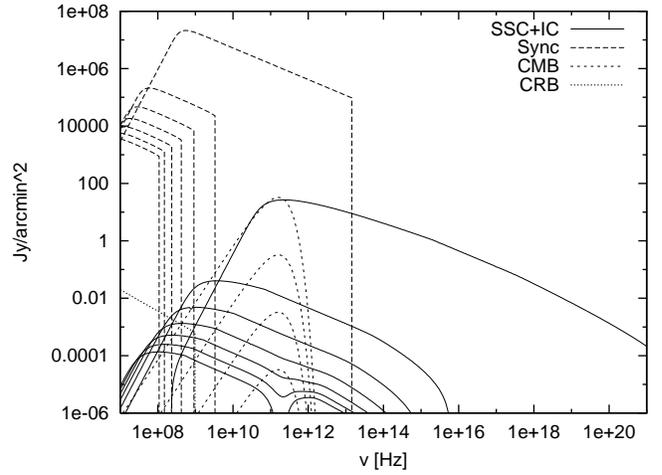,width=\figsize \textwidth,angle=0}
\end{center}
\caption[]{\label{fig:gps} Central brightness of our GPS
model. Although the surface brightness appears to be very high, the
typical $10^{-1}...10^{-2}$ arcsec sizes of GPS imply a few Jy
apparent luminosity at their spectral peak. The spectra are displayed
for a fresh GPS (top), and for possible later stages 0.017, 0.033,
0.050, 0.067, 0.083, and 0.1 Myr after a (hypothetical) jet-power
shutdown. For further
explanations see Fig. \ref{fig:CygAexp}. [Model parameters: $R =
0.05$ kpc, $B = 1 \,$mG, $p_1 = 10$, $p_2 = 3\cdot 10^4$, $s = 2.1$,
$C = 0.3\,{\rm cm^{-3}}$, $b = 1.5$, $t_0= 0.003$ Myr].}
\end{figure}

In this figure possible later stages in the evolution of the radio
source, after the supply with fresh electrons was shut down, are also
shown. In order to model these stages we assume that the radio cocoons
will expand adiabatically while they rise buoyantly in the cluster
atmosphere \cite{2001ApJ...554..261C,2001MNRAS.325..676B}. In our illustrating
example we assume that the jet supply of fresh radio plasma shuts down
at the present evolutionary stage of Cygnus A. The two radio cocoons
are assumed to expand with $b = 3/2$ and $t_0= 75$ Myr (see
Eq. \ref{eq:expansion}). This should serve as a rough model for a rise
with half sonic velocity in a singular isothermal cluster atmosphere
(cluster density $\propto$ cluster radius$^{-2}$).  We note that the evolution of
the electron spectra is not corrected for the reduced synchrotron
cooling at self absorbed frequencies. An investigation of this more
complicated problem was done by Rees \cite*{1967MNRAS.136..279R}

It is obvious, that a radio cocoon of a powerful, compact radio galaxy
produces a detectable SSC flux for a long time after the the
synchrotron radio emission vanished from the radio observable
frequency band. At low radio frequencies the SSC flux remains even for
a few Gyr, due to the slow energy losses by self-absorbed synchrotron
emission. But adiabatic losses due the expansion of the source are
able to reduce the SSC flux substantially.

The typical angular area is of the order of arcmin$^2$. Therefore
all of the instruments listed in Tab. \ref{tab:sensitivity} should be
able to detect a Cygnus-A like radio cocoon. All but LOFAR and GMRT see it best
shortly after the decay of the synchrotron emission. The latter is
because at below GHz frequencies the source emission is rising with time
during the first few 100 Myr. PLANCK and HERSCHEL should give only
relative week detection of early stages. However, if such a very
powerful radio ghost would be much closer, e.g. located in the Virgo
cluster, even later stages may be detectable with these instruments.

\subsection{GHz Peak Sources\label{sec:GPS}}

GPS are believed to be the very early stage of radio galaxies, in
which a small ($<$ kpc) radio cocoon is working its way through the
dense interstellar medium of a radio galaxy \cite[and references
therein]{2000MNRAS.319..445S,2000A&A...354..467D}. The spectral peak
at typically a GHz is likely due to synchrotron-self-absorption at
lower frequencies. Although it turns out that SSC emission of GPS is
very weak and practically unobservable, we briefly discuss such
sources for completeness since they are another extreme case of radio
galaxies.

The SSC flux of GPS should exhibit a peak around 100 GHz $(p_1/10)^2$,
where $p_1$ is the minimum electron momentum of the population. A
detection of such a peak would give deep insight into the low energy
end of the relativistic electron population. Unfortunately, the
synchrotron flux will likely also dominate this spectral regime,
unless there is a relatively low high energy cutoff in the electron
spectrum. Even in that case the brightness will be only of the order
of the CMB, which gives a very low signal for these small-scale
sources.

As an illustration, we have chosen the following parameters of our
model GPS: $R= 0.05$ kpc, $B = 1$mG, $C = 0.3\,{\rm
cm^{-3}}$, $s = 2.1$, $p_1 = 10$, and $p_2 = 3\cdot 10^4$. This
source peaks around 0.3 GHz and has a GHz luminosity of $1.4\cdot
10^{33}\,{\rm erg\,s^{-1}\,Hz^{-1}}$, as it is typical for GPS
sources.  The synchrotron and SSC spectra are shown in
Fig. \ref{fig:gps}, together with spectra for hypothetical later
stages, after a shutdown of the central engine. We assumed for this
phase $t_0 = 0.03$ Myr, and $b = 1.5$ (see Eq. \ref{eq:expansion}).

The SSC flux can easily extend into the gamma ray regime, but the
total luminosity is very low with $\nu L_\nu \sim 10^{34...35}\, {\rm
erg\,s^{-1}}$ in the X-ray to gamma ray regime. Thus, GPS are
not expected to be easily detectable via their SSC flux.

\subsection{Giant Radio Galaxies\label{sec:grg}}

\begin{figure}[t]
\begin{center}
\psfig{figure=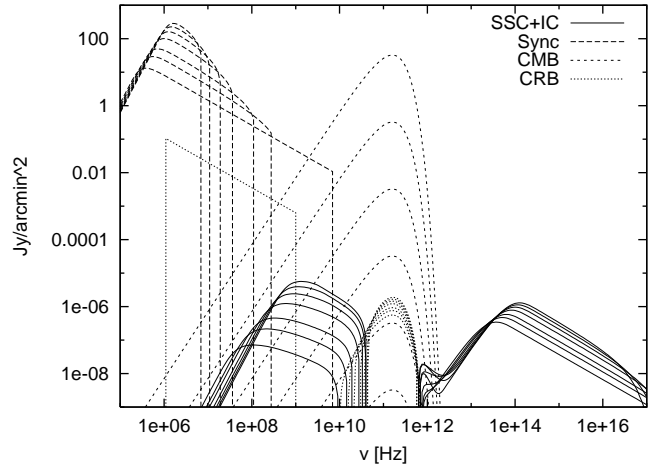,width=\figsize \textwidth,angle=0}
\end{center}
\caption[]{\label{fig:grg} Central surface brightness of our giant
radio galaxy radio cocoon. Here a compression scenario was assumed, so
that the radio luminosity at low frequencies, and the IC \& SSC flux
at most frequencies increases with time. Spectra are displayed for the
jet shutdown time and  0.3, 0.6, 1.2, 1.8, 2.4, and 3
Gyr after this.  For further explanations see
Fig. \ref{fig:CygAexp}. [Model parameters: $R= 400$ kpc, $B =
1\,\mu$G, $C = 5\cdot 10^{-7}\,{\rm cm^{-3}}$, $s = 2.5$, $p_1 = 10$,
$p_2 = 3\cdot 10^4$, $b = -1$, $t_0 = 500$ Myr].}
\end{figure}

Giant radio galaxies are an other extreme class of powerful radio
galaxies. Their radio cocoons can have diameters of a Mpc, since they
expand into a low density environment. Therefore, the magnetic and
electron energy densities are expected to be much lower in these
sources \cite[for recent reviews]{1997A&AS..123..423M,2000A&AS..146..293S}.

Here, we adopt a radius of 400 kpc, a magnetic field strength of
$1\,\mu$G, a single power law electron population with spectral index
$s = 2.5$, and a normalization of $C = 5\cdot 10^{-7}\,{\rm
cm^{-3}}$. As in the case of Cygnus A, we assume $p_{1} =10$ and
$p_{2} = 3\cdot 10^4$. Since the cooling is CMB-IC dominated, optical
thickness of the synchrotron radiation does not change these cooling
ages. The magnetic and electronic energies of this cocoon are $3\cdot
10^{59}$ erg and $2\cdot 10^{60}$ erg. The luminosity of the cocoon is
$2\cdot 10^{32}\,{\rm erg\,s^{-1}Hz^{-1}}$ at 1 GHz, corresponding to
a flux of 18 Jy at 1 GHz and 100 Mpc luminosity distance.

The radio emission  typically disappears on a 100 Myr time-scale after
the jet-power is shut down, but the IC and SSC flux remain for
Gyrs. Expansion of the fossil radio cocoon would nearly completely
diminish these already weak fluxes. But compression can strongly
increase the SSC flux even at very late stages. This is shown in
Fig. \ref{fig:grg}, where a compression, described by $b = -1$ and
$t_0 = 500$ Myr (see Eq. \ref{eq:expansion}), was assumed. Within the
displayed 3 Gyr evolution a compression of a factor $\tilde{C} = 7$ is
reached. Although the compression here is artificially introduced, it
should illustrate the consequences of the growth of structures like
groups and filaments of galaxies in which many fossils of giant radio
galaxies are expected to be embedded.

The expected angular area of a giant radio galaxy ghost is of the
order of 300 arcmin$^2$. However, only the GMRT, EVLA, and ATA
telescopes (out of the sample listed in Tab. \ref{tab:sensitivity}) at
GHz frequencies have reasonable chances to detect such a source. And
this only if it is either environmentally compressed or if
extremly long integration times are used ($10^2-10^3$ h).

\subsection{The Nearest Radio Ghost\label{sec:ourghost}}

\begin{figure}[t]
\begin{center}
\psfig{figure=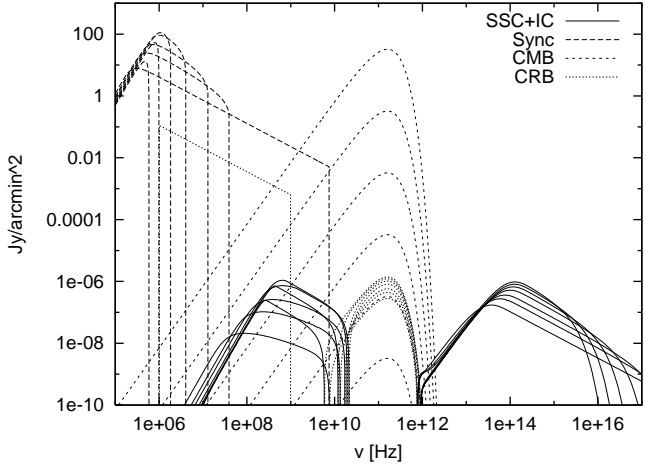,width=\figsize \textwidth,angle=0}
\end{center}
\caption[]{\label{fig:mw} Central surface brightness of a possible
radio ghost produced by our own galaxy. Also here a compression
scenario was assumed, leading to a compression by a factor of 11 in 10
Gyr), so that the radio luminosity at low frequencies, and the IC \&
SSC flux at most frequencies increases with time until an age of 6
Gyr. After this the SSC flux declines again due to the rapidly
decreasing electron high energy spectral cutoff. Spectra are displayed
for the jet shutdown time and subsequent times of 1, 2, 4, 6, 8,
and 10 Gyr after this (ignoring CMB redshift evolution for
simplicity). For further explanations see
Fig. \ref{fig:CygAexp}. [Model parameters: $R= 200$ kpc, $B =
1\,\mu$G, $C = 5\cdot 10^{-7}\,{\rm cm^{-3}}$, $s = 2.5$, $p_1 = 10$,
$p_2 = 3\cdot 10^4$, $b = -1$, $t_0 = 1$ Gyr].}
\end{figure}

The Milky Way or the Andromeda galaxy harbor very massive central
black holes with masses of $2.6\cdot 10^{6}\,M_\odot$
\cite{1997MNRAS.291..219G} and $\sim 3\cdot 10^{7}\,M_\odot$
\cite{1999ApJ...522..772K} respectively. If the growth of these black
holes was due to feeding of gas from an accretion disk, it can be
expected that this process was accompanied by a relativistic jet
outflow. If 10 \% of the rest-mass of the accreted gas was converted
into jet-power, enough radio plasma should have been produced in order
to fill an inter-galactic volume of the order of 0.1
Mpc$^3\,(B/\mu{\rm G})^{-2}$ \cite{2001APh....16...47M} with outflows
from our galaxy, and ten times as much with outflows from
Andromeda. The term $B$ denotes here the equipartition magnetic
field strength as a measure of the internal energy density. A
moderately sized nearby radio ghost of this origin may have a radius
of 200 kpc, a magnetic field strength of $1\,\mu$G, a single power law
electron population with spectral index $s = 2.5$, and an
normalization of $C = 5\cdot 10^{-7}\,{\rm cm^{-3}}$. As in the other
cases of radio plasma, we assume $p_{1} =10$ and $p_{2} = 3\cdot
10^4$. Similar to giant radio galaxies, one gets a very low SSC
surface brightness, as displayed in Fig. \ref{fig:mw}. Synchrotron and
IC cooling is not able to remove the flux within 10 Gyr. The
brightness could increase drastically due to compression, if the
environmental pressure grew considerably in the recent past due to
accretion of matter onto the local group. If such a compressed fossil
radio cocoon were nearby it would produce a relatively big total
SSC flux due to its very large angular size ($\sim 100$ deg$^2 \sim
10^6$ arcmin$^2$). Maybe ATA and GMRT would have a chance to detect
this, if these telescopes would still be sensitive to such large
scales. A detection by PLANCK should fail only by roughly one order of
magnitude. This leaves some hope that a sensitive ground (or balloon)
based single dish telescope is able to detect the large scale CMB
decrement of the order of $10^{-8}-10^{-7}$ caused by such a nearby
radio ghost.

\begin{figure}[t]
\begin{center}
\psfig{figure=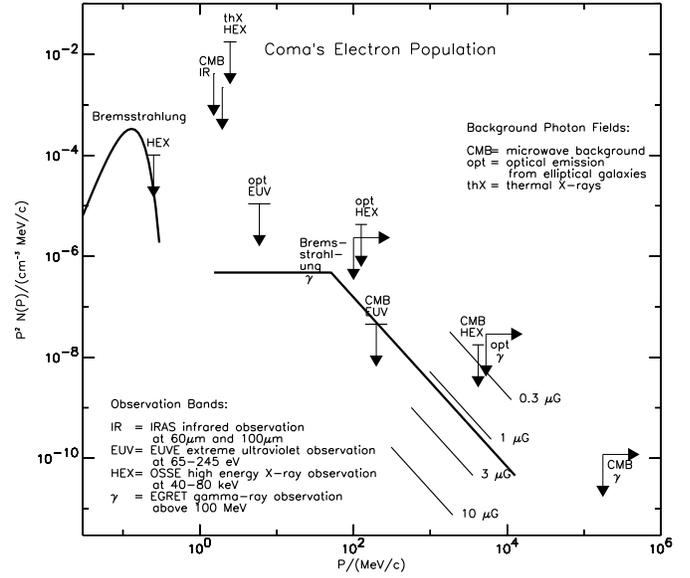,width=\figsize \textwidth,angle=0}
\end{center}
\caption[]{\label{fig:ecoma} The (central) electron spectrum in the
Coma cluster of galaxies. The thick line is the thermal distribution
and the assumed relativistic population. The upper limits result from
observational upper limits to bremsstrahlung or IC scattering fluxes
in various wave-bands. Two of them correspond to actual detections:
the extreme ultraviolet (EUV) \cite{1996Sci...274.1335L} and the high
energy X-ray (HEX) excess fluxes \cite{1999ApJ...513L..21F}. Some of
the corresponding upper limits (labeled by EUV and HEX) on the
electron spectra could therefore be data points. The thin lines give
the required electron populations in order to produce the observed
radio halo of Coma with the labeled magnetic field strengths. For
details see En{\ss}lin \& Biermann
\protect{\cite*{1998AA...330...90E}}.}
\end{figure}

\begin{figure}[t]
\begin{center}
\psfig{figure=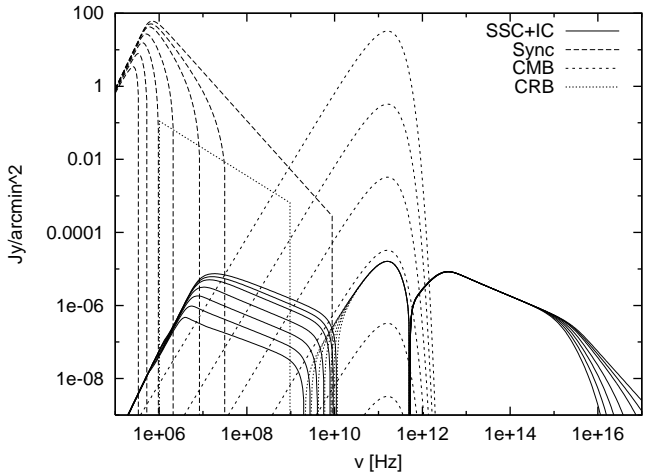,width=\figsize \textwidth,angle=0}
\end{center}
\caption[]{\label{fig:coma} Central surface brightness of the
Coma-like galaxy cluster for the electron spectrum displayed in
Fig. \ref{fig:ecoma}. Also displayed are spectra 1, 2, 3, 4, 6., 8.,
and 10 Gyr after the supply of fresh electrons stopped (ignoring CMB
redshift evolution for simplicity). They are presented only as
illustrations, since no evolution of the cluster is assumed for this
period and Coulomb losses, which would quickly remove the low energy
electrons, are also neglected. For further explanations of this figure
see Fig. \ref{fig:CygAexp}. [Model parameters: $R= 400$ kpc, $B =
1.2\,\mu$G, $C_1 = 9.34\cdot 10^{-7}\,{\rm cm^{-3}}$, $s_1 = 2$,
$p_{1,1} = 3$, $p_{1,2} = 100$, $C_2 = 2.14\cdot 10^{-3}\,{\rm
cm^{-3}}$, $s_2 = 3.68$, $p_{2,1} = 100$, $p_{2,2} = 3\cdot 10^4$].}
\end{figure}

\subsection{Galaxy Clusters\label{sec:clogl}}

Several clusters of galaxies host cluster wide diffuse radio emission,
the {\it `cluster radio halos'}. The origin of the relativistic
electrons producing the halos is not yet settled. They might be
transient features or persistent. We model a Coma cluster type radio
halo and follow its evolution after a hypothetical shut down of the
supply of fresh relativistic electrons.

We approximate the cluster by a homogeneous sphere of 400 kpc
radius. The field strength is assumed to be 1.2 $\mu$G, allowing the
same population of electrons to produce the radio emission and the
observed extreme ultra-violet excess by IC scattering of the
CMB.\footnote{In contrast to this, if the observed high energy
X-ray excess of Coma would be due to IC scattered CMB photons a
field strength of 0.2 $\mu$G would follow. The field strength is
fixed (in this case) because the responsible electrons would also be
observable by their synchrotron emission, which would form the
observed radio halo of Coma.}  The electron spectrum is described by
a broken power-law: for $3<p<100$ we use $s_1 = 2$ and $C_1 =
9.34\cdot 10^{-7}\,{\rm cm}^{-3}$, and for $100<p<3\cdot 10^4$ we use
$s_2 = 3.68$ and $C_2 = 2.14\cdot 10^{-3}\,{\rm cm}^{-3}$.

The ratio of relativistic electron to magnetic field energy density is
of the order of 50. It is dominated by the low energy electrons, which
are usually neglected in the calculation of equipartition magnetic
fields from radio observations. The ratio of relativistic to thermal
electron energy density in this model is moderate: $2.5\%$.

A compilation of the available information on the electron spectrum in
Coma was given by En{\ss}lin \& Biermann
\cite*{1998AA...330...90E}. Although a beta-profile was used in this
work to model the electron and magnetic field distribution, our
homogeneous sphere should give comparable emission for the same
densities, sufficiently accurate for our purposes. We can therefore
compare directly the above assumed electron
spectrum with the observational limits resulting from various
radiation processes in Fig. \ref{fig:ecoma}.

The spectrum of our model of the diffuse emission in Coma is shown in
Fig. \ref{fig:coma}. The synchrotron/IC cooled spectra are also shown
at times after the supply of high energy electrons has stopped.  The
low energy electron population is assumed to be preserved against
Coulomb losses by re-acceleration or injection of electrons.  This
demonstrates that some weak SSC flux could come from clusters
which do not exhibit observable synchrotron emission at all.

It should be noted that Coma is not at the upper end of the cluster
radio halo luminosity distribution. The most luminous radio halo known
to date in the cluster 1E~0657-56 is roughly 10 times brighter than
Coma \cite{2000ApJ...544..686L}. The expected SSC luminosity of this
cluster is therefore expected to be more than an order of magnitude
higher than in our model for Coma.

Coma's relativistic electron population cover roughly 300
arcmin$^2$ of the sky. This may allow detection by LOFAR of an remnant
of a radio halo in a nearby (Coma-like) cluster with very long
integration times ($10^2-10^3$~h). But GMRT, ATA, and EVLA have very
good chances of an detection at their lowest frequencies (0.1 GHz for
GMRT, GHz for ATA and EVLA) of such an aged radio halo. Experiments in
the CMB range should also come close to detection of the relativistic
CMB decrement produced by fresh or aged radio halo electron
populations, if they do not completely resolve out the source
structure. Unfortunately, any detected signal is weak compared to the
thermal Sunyaev-Zeldovich effect. In optimistic scenarios, a marginal
discrimination of such a relativistic CMB decrement may still be
possible even with PLANCK \cite{2000A&A...360..417E}.

\section{Conclusion\label{sec:concl}}

We have modeled SSC and IC spectra of different astrophysical sources
under consideration of synchrotron self-absorption, electron aging by
synchrotron, and IC losses, and adiabatic losses and gains. Although
the SSC luminosity of a typical radio source is weak compared to its
synchrotron luminosity, the SSC lifetime exceeds the synchrotron
lifetime of the source by a large margin. This property can make the
SSC emission an important probe of the fossil stages of former radio
luminous galaxies. It allows further to probe the spectral shape of
the low energy population of relativistic electrons of a source, which
is otherwise not accessible. Since the SSC flux depends strongly on
the energy density of the relativistic electrons and magnetic fields,
it is a very sensitive probe of the compression state of the fossil
radio plasma and therefore of its environment.

Powerful radio galaxies are expected to have the strongest SSC
luminosities. Their fossil radio plasma clouds (radio ghosts) can be
detectable even if adiabatic expansion is affecting them. Possible
candidates for the detection are the recently discovered cavities in
the X-ray gas of some clusters of galaxies. These are likely radio
ghosts produced by the central cluster galaxy, which are currently
on their buoyant rise through the cluster atmosphere. Examples are
listed in Sect. \ref{sec:radioghosts}.  The best time to detect the
SSC flux of a rising radio plasma bubble is during the first 100 Myr,
before adiabatic expansion has diminished its luminosity.

Giant radio galaxies, which have very extended radio cocoons, and
which are residing in low density environments, produce only a very
low SSC surface brightness. Environmental compression of such a lobe
would certainly help to make it detectable, even after a couple of
Gyrs of lurking. 

GHz peak sources have extreme SSC surface brightnesses, but very low
luminosities due to their small angular extent. If at all, their SSC
flux might be best detectable at X-ray or even gamma ray energies.

The nearest radio ghost may be produced by an earlier activity of the
Milky Way's or the Andromeda galaxy's central black hole. Although its
SSC surface brightness is expected to be very low, its large angular
extent may give us a chance to detect it with single dish
experiments at its expected emission peak in the range of 0.1-10 GHz
or at its CMB decrement.

A Coma-like galaxy cluster is expected to produce SSC flux even a long
time after a shut down of the process, which supplied the radio halo
emitting electrons.

The detection of the SSC emission of one of these classes of sources
is an observational challenge. It would be rewarded by insight into a
formerly unseen electron population and into the fossil Universe of
relativistic particle populations.  Due to its broad frequency
spectra, it can be probed with several future high sensitivity
instruments, ranging from lowest radio frequencies GMRT and LOFAR,
over microwave spacecraft MAP and PLANCK, balloon and ground based CMB
experiments, and sub-mm/IR projects ALMA and HERSCHEL satellite,
eventually even up into the X-ray and gamma-ray range.  A
multi-frequency sky survey, as will be provided by the Planck
experiment, should allow to search for the SSC and relativistic IC
spectral signature of many nearby clusters of galaxies and radio
galaxies at least in a statistical way. In addition to this, there
should be targeted observations of promising candidates, as e.g. the
recently reported X-ray cluster cavities. We demonstrated that new and
upcoming radio arrays with large collecting areas in the 0.1-1 GHz
range as LOFAR, GMRT, EVLA and ATA have excellent detection
chances.  We hope that this work stimulates observational efforts to
exploit the spectral landscape of the {\it terra incognita} of 1-100
MeV electrons residing in the intergalactic space via their combined
SSC and CMB-IC emission.

The expected long-living SSC source population of fossil radio
galaxies, and their possible impact on CMB experiments will be
investigated in a separate forthcoming paper.

\begin{acknowledgements}
It is a pleasure to thank Dan Harris, the referee, and Sebastian Heinz
for many helpful comments on the manuscript.  This research was done
with support and in the context of the {\it CMBNet} and {\it The
Intergalactic Medium} Research and Training Networks of the European
Community. It has made use of NASA's Astrophysics Data System Abstract
Service.
\end{acknowledgements}

\appendix

\section{Cooling Electron Spectra\label{sec:ecool}}

We assume that the emission region is a homogeneous sphere with radius
$R$. The magnetic fields are assumed to have isotropic
orientations. Several electron populations with isotropic momentum
distributions and power-law momentum spectra are assumed to exist
simultaneously in the emission region. Initially, they are assumed to be
power-law distributions:
\begin{equation}
\label{eq:finitial}
f_{i} (p)\, dp\,dV = C_i\,
p^{-s_i}\,\Theta(p-p_{1,i})\,\Theta(p_{2,i}-p) \,dp\,dV
\end{equation}
The index $i$ labels the different spectra, which have normalizations
of $C_i$, and spectral indices $s_i$. $\Theta(x)$ is the Heaviside
step-function, so that for dimensionless electron momenta $p = P_\e
/(m_\e \,c)$ outside $p_{1,i}<p<p_{2,i}$ the electron spectra vanish.
The freedom to have several populations allows to model complex
spectra with broken power-laws or spectral discontinuities. We also
allow that each electron population is located within a different
magnetic field strength $B_i$, so that small-scale multi-phase media
can be modeled.

After the supply of fresh electrons has stopped, the electron spectra
evolve according to synchrotron, IC, and adiabatic energy losses:
\begin{equation}
\label{eq:ploss}
- \frac{dp}{dt} = a_{0} \, (u_B + u_C) \,p^2+ 
\frac{1}{3}\,
\frac{1}{V} \,\frac{dV}{dt}\, p \,,
\end{equation}
where $a_{0} = \frac{4}{3}\, \st/(m_\e\, c)$, $V$ is the
source volume, $u_B$ and $u_C$ are the magnetic and CMB energy
densities respectively. We do not consider bremsstrahlung and Coulomb
losses in view of the very low particle density in most of the
considered sources. We assume that the expansion of the source volume
can be described by a power-law in time:
\begin{equation}
\label{eq:expansion}
V(t) = V_{0} \,\left({t}/{t_{0}} \right)^{b} = V_0/\tilde{C}(t) \,,
\end{equation}
where $\tilde{C}(t)$ denotes the compression factor.

In this case the evolution of the momentum of an electron residing in
a magnetic field strength $B_i(t) = B_i\,\tilde{C}^{2/3}$ can be given
analytically \cite[e.g.]{1997MNRAS.292..723K,2001A&A...366...26E}:
\begin{equation}
p_i(p_0, t) = \frac{p_{0}}{\tilde{C}(t)^{-\frac{1}{3}} + p_{0} \, q_i(t)}\,,
\end{equation}
where the initial momentum is $p_0$ and $q_i(t)$ can be read as the
inverse of the maximal possible electron momentum $p_{i}^*(t)$ in
population $i$:
\begin{eqnarray}
q_i(t) &=& \frac{1}{p_{i}^*(t)} = a_{0} \int_{t_{0}}^t
\!\!\!\! dt' \,(u_{B i}(t')+u_C(t')) \left(
\frac{{\tilde{C}}(t')}{{\tilde{C}}(t)} \right)^{\frac{1}{3}}\,.  \\
&=& a_{0} \, t \, \left(
\frac{\tilde{C}^{\frac{4}{3}} - \tilde{C}^{\frac{1}{b} -
\frac{1}{3}}}{1- 5\,b /3} \,u_{B{i}} + \frac{1 -
\tilde{C}^{\frac{1}{b} -
\frac{1}{3}}}{1- b /3}\, u_{C} \right)
\end{eqnarray}
($\tilde{C} = \tilde{C}(t)$ for brevity).  The electron spectra evolve
according to
\begin{eqnarray}
\label{eq:spec2}
f_i (p,t) &=& C_i \, \tilde{C}(t)^{\frac{s_i +2}{3}}\, p^{-s_i}\, \left(
1 - p\, q_{i}(t) \right)^{s_i -2}\times\nonumber\\
&& \Theta(p-p_{1,i}(t))\,\Theta(p_{2,i}(t)-p) \,,
\end{eqnarray}
where $p_{1,i}(t)= p_i(p_{1,i},t)$ and $p_{2,i}(t) =
p_i(p_{2,i},t)$. We also write $f(p)$ for $f(p,t)=\sum_i \,f_i(p,t)$
for brevity in the following and also drop the explicit time
dependence in our notation of the resulting photon spectra.

Similar evolution formulae can be derived for any arbitrary
initial electron spectrum. Since it is used in our examples, we give
here the definition of a relativistic thermal distribution:
\begin{equation}
\label{eq:fth}
f_{\e,\th}(p) = \frac{n_{\e,\th}\,\beta_{\rm th}}{K_2(\beta_{\rm th})}\, \,
p^2 \,\exp (-\beta_{\rm th}\,\sqrt{1+p^2})\,,
\end{equation}
with $K_\nu$ denoting the modified Bessel function, $n_{\e,\th}$ the thermal electron number density,
$\beta_{\rm th} = \me\, c^2/kT_\e$ the normalized thermal
beta-parameter, and $T_\e$ the electron temperature.

\section{Synchrotron Emission and Absorption\label{sec:sync}}
The pitch-angle averaged synchrotron energy losses of an
ultra-relativistic electron located in magnetic fields of strength $B$
\begin{equation}
\dot{E}_\sync(p,B) = - \frac{4}{3} \, \st\,c\,\frac{B^2}{8\,\pi}\, p^2  
\end{equation}
are radiated in the form of photons at the characteristic synchrotron frequency
\begin{equation}
\nu_{\sync}(p,B) = \frac{3\, e\, B \,p^2}{2\,\pi\,\me\,c} = \Lambda\,B\,p^2\;.
\end{equation}
The term $\Lambda = 3\, e/(2\,\pi\,\me\,c)$ is introduced for
convenience. The emissivity of a single
electron located in field strength $B_i$ is therefore
\begin{equation}
\label{eq:Psync}
P_{\sync,i}(\nu,p,t)\,d\nu = -\dot{E}_\sync (p,B_i(t)) \, \delta(\nu -
\nu_{\sync}(p,B_i(t)))\,d\nu
\end{equation}
in the monochromatic approximation, with $\delta(x)$ denoting Dirac's
delta function.  The source density of synchrotron photons
produced in the frequency interval $d\nu$ by our electron
populations is
\begin{eqnarray}
\label{eq:qsync0}
q_\sync(\nu) &=&  \sum_i \,\int_0^\infty
\!\!\! dp\,P_{\sync,i}(\nu,p,t)\,f_i(p,t)/(h\,\nu)\\
&=& \label{eq:qsync1}
\frac{\st\,c}{12\,\pi\, h\,\nu\,\Lambda} \,
\sum_i \,B_i(t) \,p \, f_i(p,t)\,|_{p=\sqrt{{\nu}/{(\Lambda\,B_i(t))}}}
\\
\label{eq:qsync3}
 &=& \sum_i \frac{\st\,c\, C_i}{12\,\pi\,h\,\Lambda^{2}}\,\tilde{C}^{-\frac{s_i+1}{3}}
\left(\frac{\nu}{\Lambda\,B_i(t)}\right)^{-\frac{s_i +1}{2}}\times \\
\nonumber  && \!\! \left( 1- q_i(t)\sqrt{\frac{\nu}{\Lambda\,B_i(t)}}
\right)^{s_i-2} \!\!\!\!\!\!\!\!\!\!\!\!
\Theta(\nu - \nu_{1,i}(t)) \Theta(\nu_{2,i}(t)-\nu).\\
\label{eq:qsync2}
&=& \sum_i \frac{\st\,c\,
C_i}{12\,\pi\,h\,\Lambda^{2}}
\left(\frac{\nu}{\Lambda\,B_i}\right)^{-\frac{s_i +1}{2}}
\!\!\!\!\!\!\!\!\!\!\!\!\!\!\!\!  \Theta(\nu - \nu_{1,i})
\Theta(\nu_{2,i}-\nu).
\end{eqnarray}
Eq. \ref{eq:qsync1} is the general form and Eq. \ref{eq:qsync2} is
valid for the power-law electron populations given in
Eq. \ref{eq:finitial}.  Here $\nu_{1,i}(t) = \Lambda\,B_i(t)\,p_{1,i}^2(t)$ and
$\nu_{2,i}(t) = \Lambda\,B_i(t)\,p_{2,i}^2(t)$ are the lower and upper frequency
cutoffs.

The synchrotron absorption coefficient within the the monochromatic
approximation (Eq. \ref{eq:Psync}) can be calculated from the generic
form given in Rybicki \& Lightman \cite*{1979rpa..book.....R}:
\begin{eqnarray}
\label{eq:alphanu0}
&&\!\!\!\!\alpha_\nu \!= \!- \frac{c^2}{8\,\pi\,\nu^2\,m_\e\,c^2} \sum_i \,\int_0^\infty
\!\!\! dp\,P_{\sync,i}(\nu,p)\,p^2 \frac{\partial}{\partial p}
\frac{f_{i}(p)}{p^2} \\
\label{eq:alphanu1}
&&\!\!\!\!= \!\sum_i \frac{(2+s_i-4\,p\,q_i(t))\,\st
\,c}{96\,\pi^2\,m_\e\,\Lambda^3\,B_i(t)}\,
\frac{f_i(p,t)\, p^{-4}}{1-p\,q_i(t)}\,|_{p=\sqrt{\frac{\nu}{\Lambda\,B_i(t)}}}\\
\label{eq:alphanu2}
&&\!\!\!\!= \!\sum_i \frac{(2+s_i)\,\st \,c\,C_i}{96\,\pi^2\,m_\e\,\Lambda^3\,B_i}\! 
\!\left( \frac{\nu}{\Lambda\,B_i}
\right)^{-\frac{s_i+4}{2}}\!\!\!\!\!\!\!\!\!\!\!\!\!\!\Theta(\nu - \nu_{1,i}) \,
\Theta(\nu_{2,i}-\nu)
\end{eqnarray}
Spectral edge effects are ignored here. Eq. \ref{eq:alphanu2} again
refers to the pure power-law spectra case.  We define the optical
depth of the source by $\tau(\nu) = R\, \alpha_\nu$.

The synchrotron surface brightness  $S_\sync$ is the integrated
emissivity along the line of sight through the source, corrected for
the synchrotron self-absorption.  For a line-of-sight passing the
source center at distance $r$, this is:
\begin{eqnarray}
S_\sync(\nu,r) &=& \frac{h\,\nu}{4\,\pi} \int_{-\sqrt{R^2-r^2}}^{\sqrt{R^2-r^2}}
\!\!\!\!\!\! \!\!\!\!\!\! \!\!\!\!\!\! dz \, q_\sync(\nu,\sqrt{z^2+r^2})\,
e^{-\alpha_\nu\,(\sqrt{R^2-r^2}-z)}\\ &=&
\frac{q_\sync(\nu)\,h\,\nu\,R}{4\,\pi\,\tau(\nu)}
\left(1-e^{-2\,\alpha_\nu \sqrt{R^2-r^2}}\right)
\end{eqnarray}
From this the total synchrotron source luminosity can be obtained via
surface integration:
\begin{eqnarray}
\label{eq:Lsync}
L_\sync(\nu) &=& \frac{q_\sync(\nu)\,h\,\nu\,\pi R^3}{2\,\tau^3}
\left[(1+2\tau) e^{-2\tau} - (1-2\tau)\right]\\
\label{eq:Lsync1}
&\approx& \left\{ 
\begin{array}{ll}
4\,\pi\,q_\sync(\nu)\,h\,\nu\,R^3/3 & \mbox{for}\, \tau(\nu)
\rightarrow 0 \\
\label{eq:Lsync2}
\pi\,q_\sync(\nu)\,h\,\nu\,R^2/\alpha_\nu & \mbox{for}\, \tau(\nu)
\rightarrow \infty
\end{array}\right.
\end{eqnarray}

The synchrotron photon number density as a function of position and
frequency is best calculated with the help of the radiative Greens
function in a homogeneously absorbing medium, which is
\begin{equation}
G_\nu (\vec{r}',\vec{r}) = \frac{e^{-\alpha_\nu\, |\vec{r}'
-\vec{r}|}}{4\,\pi\,c\, |\vec{r}' -\vec{r}|^2}\,.
\end{equation}
The synchrotron photon density within the source (for $r<R$) is given
by an integration over the spherical emission volume $V$:
\begin{eqnarray}
&&\!\!\!\!n_\sync (\nu,r)= \int_V dr'^3\, q_\sync(\nu,\vec{r}')\,G_\nu (\vec{r}',\vec{r})\\
&&= \frac{q_\sync(\nu) }{4\,c \,\alpha_\nu^2\, r} \left[
4 \,\alpha_\nu\, r+ 
\right. \\
\label{eq:nsync}
&&\;\;\;\;\left.  \alpha_\nu^2\,(R^2-r^2)\,({\rm Ei}(-\alpha_\nu (R+r))- {\rm Ei}(-\alpha_\nu
(R-r))) -\right.\nonumber\\
&& \;\;\;\;\left.  2\,e^{-\alpha_\nu R} \left( (1 + \alpha_\nu
\,R)\sinh(\alpha_\nu \,r) + \alpha_\nu\,r\cosh(\alpha_\nu \,r) \right)\right]\nonumber\,,
\end{eqnarray}
where ${\rm Ei}(x)$ denotes the exponential integral.  The total
number of synchrotron photons within the source is then given by
\begin{eqnarray}
N_\sync(\nu) &=& 4\,\pi\,\int_0^R \!\!dr\,r^2\, n_\sync(\nu,r) \\
\label{eq:Nsync}
&=& \frac{q_\sync(\nu)\,\pi\,R^4}{2\,c \,\tau_\nu^4} \left[
1-2\,\tau^2 + \frac{8}{3}\,\tau^3 - (1+2\,\tau)\,e^{-2\tau}
\right]\\
&\approx& \left\{ 
\begin{array}{ll}
{q_\sync(\nu)\,\pi\,R^4}/{c} & \mbox{for}\, \tau(\nu) \rightarrow 0\\
{4\,\pi}\,{q_\sync(\nu)\,R^3}/({3\,\alpha_\nu\, c}) & \mbox{for}\, \tau(\nu) \rightarrow \infty
\end{array}
\right.
\end{eqnarray}

External photon fields do not penetrate the source completely for
frequencies in the synchrotron self-absorption frequency regime.  A
homogeneous cosmic background brightness $S_{\rm CB}(\nu)$ produces a
photon field within the source according to
\begin{equation}
n_{\rm CB}(r,\nu) = \int\,d\Omega\,
e^{-\alpha_\nu\,d(r,\Omega)}\,S_{\rm CB}(\nu)/(h\,\nu)\, ,
\end{equation}
where $d(r,\Omega)$ is the distance of a point at radius $r$ from the
source surface at radius $R$ in the direction $\Omega$. The volume and
central line-of-sight integrated number densities can be approximated
by
\begin{eqnarray}
\label{eq:NCB}
N_{\rm CB}(\nu)  &\approx& \frac{16\,\pi^2\,R^3}{3\,h\,\nu}\,S_{\rm CB}(\nu)
\frac{(1+(3\,\tau/4)^4)^{\frac{1}{4}}}{1+\tau^2}\\
\Sigma_{\rm CB}(\nu)  &\approx& \frac{8\,\pi\,R}{h\,\nu}\,S_{\rm CB}(\nu)
\frac{(1+(\tau/4)^2)^{\frac{1}{2}}}{1+\tau^2}\,.
\label{eq:SCB}
\end{eqnarray}
These approximations have an accuracy of $10\%$ and they are
asymptotically correct for $\tau = \tau(\nu) \rightarrow 0$ and $\tau
\rightarrow \infty$.

\section{Comptonization\label{sec:IC}}

A photon with frequency $\nu$ is on average shifted to the
frequency
\begin{equation}
\nu_\ic(\nu,p) = (\frac{4}{3}\,p^2+1)\,\nu \,.
\end{equation}
in an IC collision with a relativistic electrons with momentum
$p$. This relation is assumed to hold exactly in  the monochromatic
approximation. The IC photon production spectrum is therefore
\begin{equation}
q_\ic(\nu,r) \!= \!\st c \!\!\int
\!\!\!d\nu'\!\,n(\nu',r)\!\!\int\!\!\! dp\,
f(p)\,\delta(\nu-\nu_\ic(\nu',p)),
\end{equation}
where $n(\nu,r)$ is the target photon spectral density.
We write 
\begin{equation}
\nu_0(\nu,p) = \nu/(\frac{4}{3}\,p^2+1)
\end{equation}
for the target photon frequency $\nu_0$, which is required in order to
produce a scattered photon with $\nu$ by an electron with momentum
$p$. A volume integration of the IC emissivity $q_\ic(\nu,r)$ gives
the total IC flux:
\begin{equation}
\label{eq:Lic}
L_\ic (\nu) = \st\,c\,\int\,dp\,f(p)\,N(\nu_0(\nu,p))\,h\,\nu_0(\nu,p)\,.
\end{equation}
Here $N(\nu)$ is the total target photon spectrum within the source
region. Similarly, the central surface brightness is given by
\begin{equation}
\label{eq:Bic}
S_\ic (\nu) = \st\,c\,\int\,dp\,f(p)\,\Sigma(\nu_0(\nu,p))\,h\,\nu_0(\nu,p)\,,
\end{equation}
where $\Sigma(\nu)$ is the central line-of-sight integrated photon spectrum.

Some care has to be taken in the case of significant overlap of the
target photon spectrum and the spectral range of interest, as it is
the case for the CMB-IC process. The up-scattered CMB photons are
missing at CMB frequencies. Therefore a negative brightness can be
superimposed on the CMB due to IC scattering. In order to take this
into account, we use in the case of the CMB-IC process 
\begin{equation}
\label{eq:Lcmbic}
L_{\rm CMB-IC} (\nu) = \st\,c\!\!\int\!\! dp\,f(p)\,(N(\nu_0)\,h\,\nu_0-N(\nu)\,h\,\nu)
\end{equation}
and 
\begin{equation}
\label{eq:Bcmbic}
S_{\rm CMB-IC} (\nu) =
\st\,c\!\!\int\!\!dp\,f(p)\,(\Sigma(\nu_0)\,h \nu_0- \Sigma(\nu)\,h \nu)\,,
\end{equation}
where $\nu_0 = \nu_0(\nu,p)$.

In the case of the synchrotron photons being scattered (SSC process)
the central IC surface brightness of the source is:
\begin{equation}
\label{eq:Bssc}
S_\ssc(\nu) = \frac{\st \, R^2}{4\,\pi}\, \int\!\!\!dp
\;q_\sync(\nu_0)\, h \, \nu_0\,f(p)\,g(\tau(\nu_0)),
\end{equation}
where $\nu_0 = \nu_0(\nu,p)$ for brevity.
The geometric factor $g(\tau(\nu))$ is just the normalized
central-line-of-sight integrated number density of synchrotron
photons:
\begin{equation}
\label{eq:gtau}
g(\tau(\nu)) = \frac{c}{q_\sync(\nu)\,R^2}\,\int_{-R}^R \!\!\!\!\!\!\!\!dr\, n_\sync(\nu,r) 
\approx \left(\frac{g_0^\beta + (g_\infty \,\tau )^\beta}{(1 +
\tau^\beta)^2}\right)^{{1}/{\beta}}\!\!\!\!\!\!\!\!\!.
\end{equation}
This factor is approximated to better than $4\%$ accuracy by the
asymptotically exact expression (for $\tau_\nu \rightarrow 0$ and
$\tau_\nu \rightarrow \infty$) with $g_0 = (4 + \pi^2)/{8}$, $g_\infty
= 2$, and $\beta = 5/4$.

If the spectra are power-laws, if the optical depth is small over the
whole synchrotron spectral ranges, and if all IC scattering electrons
are ultra-relativistic ($p \gg 1$) Eqs. \ref{eq:Lic} and \ref{eq:Bssc}
can be integrated analytically:
\begin{eqnarray}
\label{eq:LSSC}
L_\ssc(\nu) &=& \frac{\st^2\,c\,R^4}{12\,\Lambda}
\sum_{i,j} C_i\,C_j\,B_j \left( \frac{3\,\nu}{4\,\Lambda\,B_j}
\right)^{-\frac{s_j-1}{2}}\!\!\!\!\!\!\!\!\!\!\!\kappa_{ij}\,,\\
S_\ssc(\nu) &=& \frac{\st^2\,c\,R^2\,g_0}{48\,\pi^2\,\Lambda}
\sum_{i,j} C_i\,C_j\,B_j \left( \frac{3\,\nu}{4\,\Lambda\,B_j}
\right)^{-\frac{s_j-1}{2}}\!\!\!\!\!\!\!\!\!\!\!\kappa_{ij}\,,
\end{eqnarray}
where
\begin{eqnarray}
\kappa_{ij} &=& \nonumber
%\!\!\!\!\!\!\!\!\!\!\!\!\!\!\!\!\!\!\!\!\!\!\!\!\!\!
\!\!\left\{
\begin{array}{ll}
0				\!\!&;p_{1,ij}\ge p_{2,ij}\\
\ln(p_{2,ij}/p_{1,ij})	\!\!&;p_{1,ij} <  p_{2,ij}\,\&\,s_i =
s_j\\
(p_{2,ij}^{s_j-s_i}-p_{1,ij}^{s_j-s_i})/({s_j-s_i})	\!\!&;p_{1,ij} <  p_{2,ij}\,\&\,s_i \ne s_j\\
\end{array}
\right.\\
\nonumber
p_{1,ij}\!\! &=&
\max\{ p_{1,i}, (3\,\nu/(4\,B_j\,\Lambda\,p_{2,j}^2))^{1/2}\}\\
\nonumber
p_{2,ij}\!\! &=& \min\{p_{2,i},(3\,\nu/(4\,B_j\,\Lambda\,p_{1,j}^2))^{1/2}\}\,.
\end{eqnarray}
This formula gives quite accurate results well above the synchrotron
self-absorption break in the SSC spectrum

It should be noted, that a more exact treatment of the SSC process
would require taking the anisotropy of the synchrotron radiation
field into account in the IC calculations. This could in principle be
done within the formalism described by Brunetti
\cite*{2000APh....13..107B}. But it turns out that the difference
between isotropic and anisotropic IC scattering for the line-of-sight
integrated IC flux in a spherical problem is too small to be of
importance for our rough estimates \cite{1999AA...344..409E}.

A second note: we do not correct for synchrotron self-absorption of
the inverse Compton scattered radiation for the following
reasons. Synchrotron self-absorption occurs only at frequencies at
which also 
synchrotron emission is produced. At higher (than synchrotron emission)
frequencies we therefore do not need to correct for it. At lower
frequencies the spectrum is dominated by orders of magnitude by the
synchrotron flux in all our examples and therefore the SSC emission is
unobservable at these frequencies. This justifies our optical thin
treatment of the IC photon escape.

\end{document}